\documentclass[preprintnumbers,twocolumn,showpacs,amsmath,amssymb]{revtex4}

\usepackage{epsfig}
\usepackage{graphicx}
\usepackage{dcolumn}
\usepackage{bm}

\newcommand{\be}{\begin{equation}}
\newcommand{\ee}{\end{equation}}

\newcommand{\bea}{\begin{eqnarray}}
\newcommand{\eea}{\end{eqnarray}}
\newcommand{\bfig}{\begin{figure}}
\newcommand{\efig}{\end{figure}}
\newcommand{\bc}{\begin{center}}
\newcommand{\ec}{\end{center}}

\begin{document}
\preprint{ZU-TH 03/06} 

\title{Isolated Photons in Deep Inelastic Scattering}

\author{A.\ Gehrmann-De Ridder$^a$, T.\ Gehrmann$^b$, E.\ Poulsen$^b$}
 \affiliation{$^a$ Institute for Theoretical Physics, ETH, CH-8093 Z\"urich,
Switzerland\\
$^b$ Institut f\"ur Theoretische Physik,
Universit\"at Z\"urich, CH-8057 Z\"urich, Switzerland}

\date{\today}

\begin{abstract}
Photon radiation at large transverse momenta at colliders
is a detailed probe of hard interaction dynamics. 
The isolated photon production cross section 
in deep inelastic scattering was measured recently by the 
ZEUS experiment, and found to be considerably larger than
theoretical predictions obtained with widely used event generators. To 
investigate this discrepancy, we perform a dedicated 
parton-level calculation of this observable, including
contributions from fragmentation and large-angle radiation. 
Our results are in good agreement with all aspects of the experimental 
measurement. 
\end{abstract}

\maketitle


Production of final state photons is a classical particle physics 
measurement, since its cross section 
 can be computed in principle in the well-established framework of 
Quantum Electrodynamics. In turn, 
photon production cross sections were suggested as a probe of auxiliary 
quantities in particle physics, such as the measurement of the 
gluon distribution in the proton from isolated photon production at hadron 
colliders. With the advent of precise data on this 
observable (see~\cite{ppgam} for the most recent data), it 
was realized that the theoretical description of photons produced in 
reactions also involving final-state hadrons is less straightforward than 
expected, resulting in extensive theoretical discussion. 
For isolated photon production at hadron colliders, it turns out that 
effects from 
the fragmentation of a hadronic jet into a single, highly energetic 
photon~\cite{vogelsang} and 
effects from the resummation of threshold and recoil 
corrections~\cite{laenen,catani} had to be included to obtain a satisfactory 
description of experimental data. A very sensitive issue is the 
definition of isolated photons produced in association with hadrons, since 
a completely isolated photon is not an infrared safe observable in 
quantum chromodynamics (QCD). At present, this isolation is 
usually accomplished experimentally 
by admitting only a limited amount of hadronic energy inside a cone 
around the photon direction. 

Recently, the ZEUS collaboration at DESY HERA reported a 
measurement~\cite{zeus} of 
the inclusive production cross section for isolated photons in deep 
inelastic scattering (DIS). The normalization of the experimentally 
determined cross section turned out to exceed the cross section expected 
from the multi-purpose event generator programs
HERWIG~\cite{herwig} and PYTHIA~\cite{pythia} by factors 7.9 and 
2.3 respectively. Even after normalizing the total event rate, 
none of these programs was able to describe all kinematical 
dependencies of the measured cross section. Since the same event 
generator programs are used frequently to estimate photon production 
backgrounds for new particle searches in other collider environments, 
it appears to be very important to determine the origin of these large 
discrepancies. 
In a subsequent study~\cite{mrst} the ZEUS measurement 
was analyzed in view of determining the photon distribution in the proton, 
relevant for electroweak radiative corrections at colliders.

By further analyzing the hadronic final state in 
isolated photon production in DIS, it is possible to define 
photon-plus-jet cross sections. In~\cite{zeus}, the 
isolated-photon-plus-one-jet  cross section was also measured and found 
in good agreement with the theoretical prediction~\cite{kramer}. 
This observation renders the discrepancy in the inclusive 
isolated photon cross section even more intriguing, since the 
inclusive cross section can in principle be obtained by summing all 
isolated-photon-plus-$n$-jet  cross sections, starting with $n=0$. 
To investigate the origin of the discrepancy, we performed a new calculation 
of the inclusive isolated photon cross section in DIS. 

Production of photons in deep inelastic scattering is described by the 
leading order parton-level process
$$ q(p_1) + l(p_2) \to \gamma(p_3) + l(p_4) + q(p_5)\;,$$ 
where $q$ represents a quark or anti-quark, and $l$ a lepton or 
anti-lepton. The measurable cross section for lepton-proton 
scattering $\sigma (ep \to e\gamma X)$ 
is obtained by convoluting the parton-level lepton-quark cross 
section $\hat\sigma (eq \to e\gamma q)$ 
with the quark distribution functions in the proton. 
In the scattering amplitudes for this process, the lepton-quark interaction 
is mediated by the exchange of a virtual photon, and the final state photon 
can be emitted off the lepton or the quark. Consequently, one finds three
contributions to the cross section, coming from the squared amplitudes 
for radiation off the quark ($QQ$) or the lepton ($LL$), as well as the
interference of these amplitudes ($QL$). These were computed originally as part
of the QED radiative corrections to deep inelastic scattering~\cite{riemann},
where the final state photon remains unobserved.  
The $QL$ contribution is odd under charge exchange, such that it contributes
with opposite sign to the cross sections with $l=e^-$ and $l=e^+$. 

The isolated photon rate in deep inelastic scattering is defined by imposing 
a number of kinematical cuts on the final state particles. In the ZEUS 
analysis (which combined three data samples:  38 pb$^{-1}$ $e^+p$ at 
$\sqrt{s} = 300$~GeV, 68 pb$^{-1}$ $e^+p$ at 
$\sqrt{s} = 318$~GeV and 16 pb$^{-1}$ $e^-p$ at 
$\sqrt{s} = 318$~GeV), these were chosen as follows: 
virtuality of the process, as determined from the outgoing electron 
$Q^2 = -(p_4-p_2)^2 > 35$~GeV$^2$, 
outgoing electron energy $E_e > 10$~GeV and angle 
$139.8^\circ < \theta_e < 171.8^\circ$, outgoing photon transverse energy
5~GeV$<E_{T,\gamma}<$10~GeV and rapidity $-0.7<\eta_\gamma<0.9$. Photon 
isolation from hadrons is obtained by requiring the photon to carry at 
least 90\% of the energy found in a cone of radius $R=1.0$ in the $\eta-\phi$ 
plane around the photon direction. This cone-based isolation procedure 
is commonly used to define isolated photons produced in a hadronic 
environment. A minimal amount of hadronic activity inside the cone 
has to be allowed in order to ensure infrared finiteness of the observable.
This cone-based isolation could face  
theoretical problems only if the cone size is chosen much smaller than
unity~\cite{catani}, as often required for new particle searches.
Finally, to eliminate contributions from elastic 
Compton scattering, observation 
of hadronic tracks in the ZEUS central tracking detector was required. This 
cut can not be translated directly into the parton model calculation. 
We translate this cut as follows: the ZEUS central tracking detector~\cite{ctd}
covers in the forward region rapidities $\eta < 2$. Requiring tracks in 
this region amounts to the current jet being at least partially contained 
in it. Assuming a current jet radius of one unit in rapidity, this amounts to 
a cut on the outgoing quark rapidity $\eta_q<3$, which we apply here.  
Varying this cut results only in small variations of the resulting cross 
sections.

In the $QQ$ contribution, the photon can have two possible origins: 
the direct radiation off the 
quark and the fragmentation of a hadronic jet into a
photon carrying a large fraction of the jet energy.
While the former direct process takes place
at an early stage in the process of hadronization and can be calculated
in perturbation theory, the
fragmentation contribution is primarily due to a long distance process
which cannot be calculated within perturbative methods.
The latter is described
by the process-independent quark-to-photon fragmentation
function~\cite{koller} $D_{q\to \gamma}(z)$,
 where $z$ is the momentum 
fraction carried by the photon.
$D_{q\to \gamma}(z)$ must be determined by experimental data.
Furthermore, when the photon is radiated somewhat later during the
hadronization process, in addition to this genuinely non-perturbative
fragmentation process, the emission of a photon collinear to the primary
quarks is also taking place and has to be taken into account, 
giving rise to a collinear singularity.
As physical cross sections are necessarily finite, this collinear
singularity gets factorized into the fragmentation function defined at 
some factorization scale $\mu_{F,\gamma}$.
The factorization procedure of final state collinear singularities
in fragmentation functions used here is
of the same type as the procedure used to absorb initial state collinear
singularities into the parton distribution functions~\cite{AP}. 
In the present calculation, we use the phase space slicing method~\cite{gg}
to handle the collinear quark-photon singularity, as described in 
detail in~\cite{kramer,andrew}. As a result, the parton-level cross section 
 $\hat\sigma (eq \to e\gamma q)$ contains a collinear 
divergence, which is compensated by adding the parton-level 
cross section for the fragmentation process 
 $\hat\sigma (eq \to e q)\otimes D_{q\to \gamma}(z)$ to it. Since the 
photon is required to carry at least 90\% of the energy of the quark-photon 
cluster, $D_{q\to \gamma}(z)$ is probed only for $z\geq 0.9$. 

The only data constraining $D_{q\to \gamma}(z)$ come from final state 
photon radiation in electron-positron annihilation at LEP
(some earlier evidence for the non-vanishing of  $D_{q\to \gamma}(z)$
was obtained by the EMC experiment~\cite{emc} 
in deep inelastic muon-proton scattering). 
Using the 
method described in~\cite{andrew}, the 
ALEPH collaboration has performed~\cite{aleph}
 a direct measurement of $D_{q\to \gamma}(z)$ 
from the photon-plus-one-jet rate in $e^+e^-$ 
using a leading order (LO) theoretical calculation. 
In the method of~\cite{andrew},
logarithms of  $\mu_{F,\gamma}$ are not resummed, such that any
cross sections computed with $D_{q\to \gamma}(z)$ of~\cite{aleph}
are completely independent of $\mu_{F,\gamma}$. Next-to-leading order 
(NLO) corrections to the photon-plus-one-jet rate in $e^+e^-$ are 
known~\cite{ggam}, 
and allow for a 
determination of $D_{q\to \gamma}(z)$ at NLO~\cite{ggam1} from the ALEPH data.

Several other parameterizations 
of photon fragmentation functions were proposed in the literature, 
based on models for the non-perturbative components~\cite{grv,bfg}. 
These parameterizations incorporate a resummation of logarithms of 
 $\mu_{F,\gamma}$, such that physical cross sections acquire a residual 
dependence on $\mu_{F,\gamma}$. Advantages and drawbacks of this 
resummation applied to different observables are discussed 
in~\cite{lepphoton}.
The BFG parameterizations~\cite{bfg}, yield a satisfactory
description~\cite{lepphoton} of the ALEPH data.
Furthermore, a
 measurement of the inclusive photon spectrum in hadronic $Z$ boson 
decays~\cite{opal} was
made by the OPAL collaboration. 
The OPAL data are consistent~\cite{lepphoton} with the ALEPH and 
BFG parameterizations. 

In our calculation, we use the ALEPH leading order 
parameterization~\cite{aleph} as default, and the BFG (type I)  
parameterization~\cite{bfg}, evaluated for $\mu_{F,\gamma}^2=Q^2$ 
for comparison. 
The factorization scale $\mu_F^2$ for the parton distributions is 
$Q^2$ for the $QQ$ subprocess.  

In the $LL$ subprocess
(which is the only subprocess included in~\cite{mrst}), 
the final state photon is radiated off the lepton. 
Consequently, the momentum of the final state lepton can not be used to
determine the invariant four-momentum transfer between the lepton and 
the quark, which is in this subprocess given by 
$Q_{LL}^2= -(p_5-p_1)^2$, with $Q_{LL}^2 < Q^2 = Q^2_{QQ}
= -(p_4-p_2)^2$. 
In principle, $Q_{LL}^2$ is unconstrained by the kinematical cuts, and the 
squared matrix element for the $LL$
 subprocess contains an explicit $1/Q^2_{LL}$. 
The track requirement, implemented through a cut on the 
outgoing quark rapidity, enforces a minimum $Q^2_{LL}$, thus avoiding a 
singularity in the subprocess cross section. Some care has to be taken 
in the choice of factorization scale $\mu_F^2$ in the LL subprocess. 
In a leading order parton model
calculation, $\mu_F^2$ should ideally be taken to be 
the invariant four-momentum transfer to the quark, i.e.\ $Q_{LL}^2$ for the 
$LL$ subprocess. Even applying the quark rapidity cut, 
$Q^2_{LL}$ can assume low values $Q^2_{LL}\sim \Lambda_{QCD}^2$, where the 
parton model description loses its meaning. Because of the cuts,
this kinematical region yields only a small contribution to the cross section. 
To account for it in the parton model framework, we introduce a minimal 
factorization scale $\mu_{F,{\rm min}} = 1$~GeV, and 
choose for the $LL$
subprocess $\mu_F$ = max($\mu_{F,{\rm min}}$,$Q_{LL}$), and for the 
$QL$ interference subprocess $\mu_F$ = 
max($\mu_{F,{\rm min}}$,$(Q_{LL}+Q_{QQ})/2$).  
This fixed factorization scale is 
an approximation to more elaborate procedures to extend the parton model 
to low virtualities~\cite{bk}, but sufficient in the present context.

This procedure for the scale setting in the 
$LL$ and $QL$ subprocesses is similar to what is done in the related
process of 
electroweak gauge boson production in electron-proton
collisions~\cite{bvz}. The major difference to~\cite{bvz} is that 
the cross section for isolated photon production in DIS vanishes for 
$Q^2_{QQ,LL} \to 0$, while being non-vanishing for vector boson production.
Consequently, in~\cite{bvz} the calculation of deep inelastic gauge 
boson production had to be supplemented by photoproduction of gauge 
bosons at $Q^2=0$, with a proper matching of both contributions at a
low scale. This is not necessary in our case. 
\begin{ruledtabular}
\begin{table}[t]
\begin{tabular}{lr}
 & $\sigma(ep\to e\gamma X)$ [pb] \\[1mm] \hline \\[-3mm]
ZEUS & 5.64 $\pm$ 0.58 (stat.) ${+0.47 \atop -0.72}$ (syst.) \\[1mm] \hline
Theory  & 5.39 \\[0.3mm] \hline
$e^+p$ (318~GeV) & 5.48 \\
$e^+p$ (300~GeV) & 5.32 \\ 
$e^-p$ (318~GeV) & 5.14 \\[0.3mm] \hline
QQ only & 2.87 \\
LL only & 2.39 \\
QL only & 0.13 \\[0.3mm] \hline
Theory using BFG $D_{q\to \gamma}$ & 5.51 
\end{tabular}
\caption{Values for the isolated photon cross section in deep inelastic 
scattering using the cuts of the ZEUS analysis~\protect{\cite{zeus}}. The 
theory prediction 
is the weighted 
average of electron and positron induced cross sections at different
energies
as analyzed by the experiment. By default, the ALEPH parameterization of 
the quark-to-photon fragmentation function~\cite{aleph} is used, with results 
obtained using  
the BFG parameterization~\cite{bfg} listed for comparison.
\label{tab_werte}}
\end{table}
\end{ruledtabular}

For the numerical evaluation of the cross sections, we 
use the CTEQ6L leading order parameterization~\cite{cteq} 
of parton distributions. Using the ZEUS cuts and the ZEUS composition of the 
data sample at different energies and with electrons and positrons, we 
obtain a theoretical prediction 
for the isolated photon cross section in DIS
of 5.39~pb, to be compared to the 
experimental value of 5.64$\pm$0.58(stat.)${+0.47 \atop -0.72}$(syst.).
The total cross section is therefore well reproduced by our 
calculation. 
\begin{figure}[tbh]
\epsfig{file=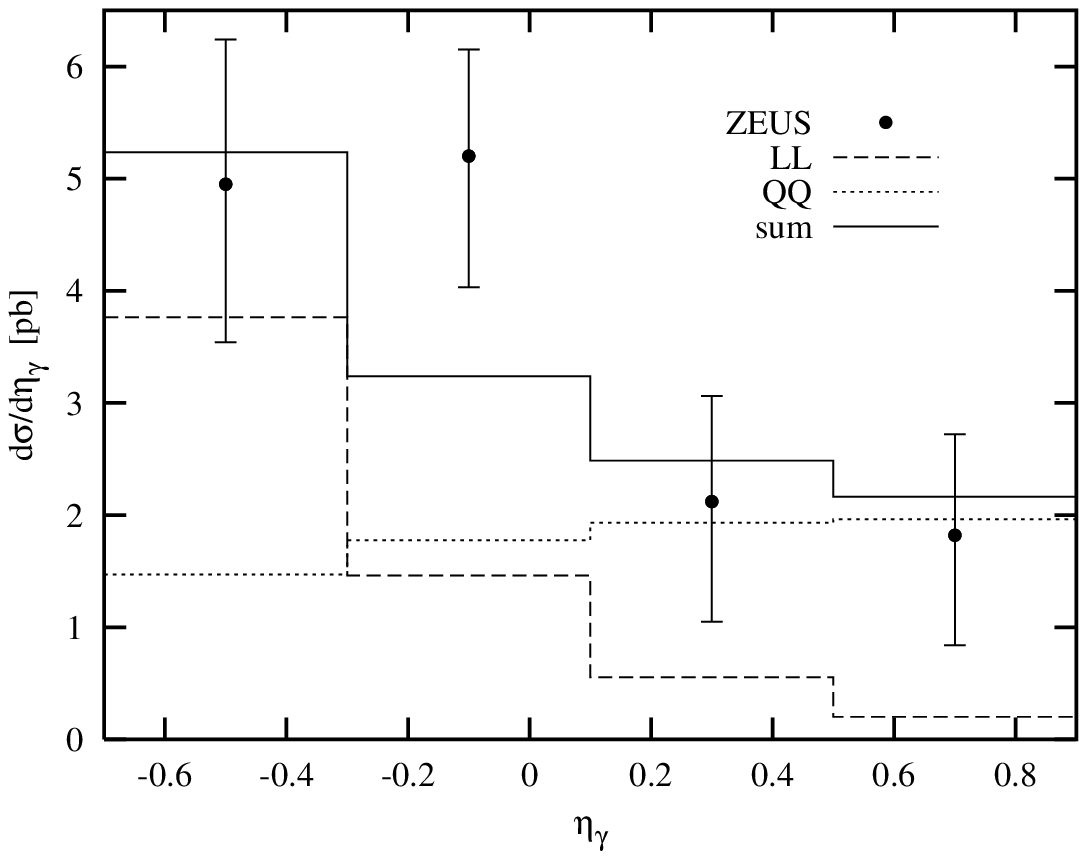,angle=-90,width=6.1cm} 
\caption{Rapidity distribution of isolated photons, 
compared to ZEUS data.\label{fig:rap}}
\epsfig{file=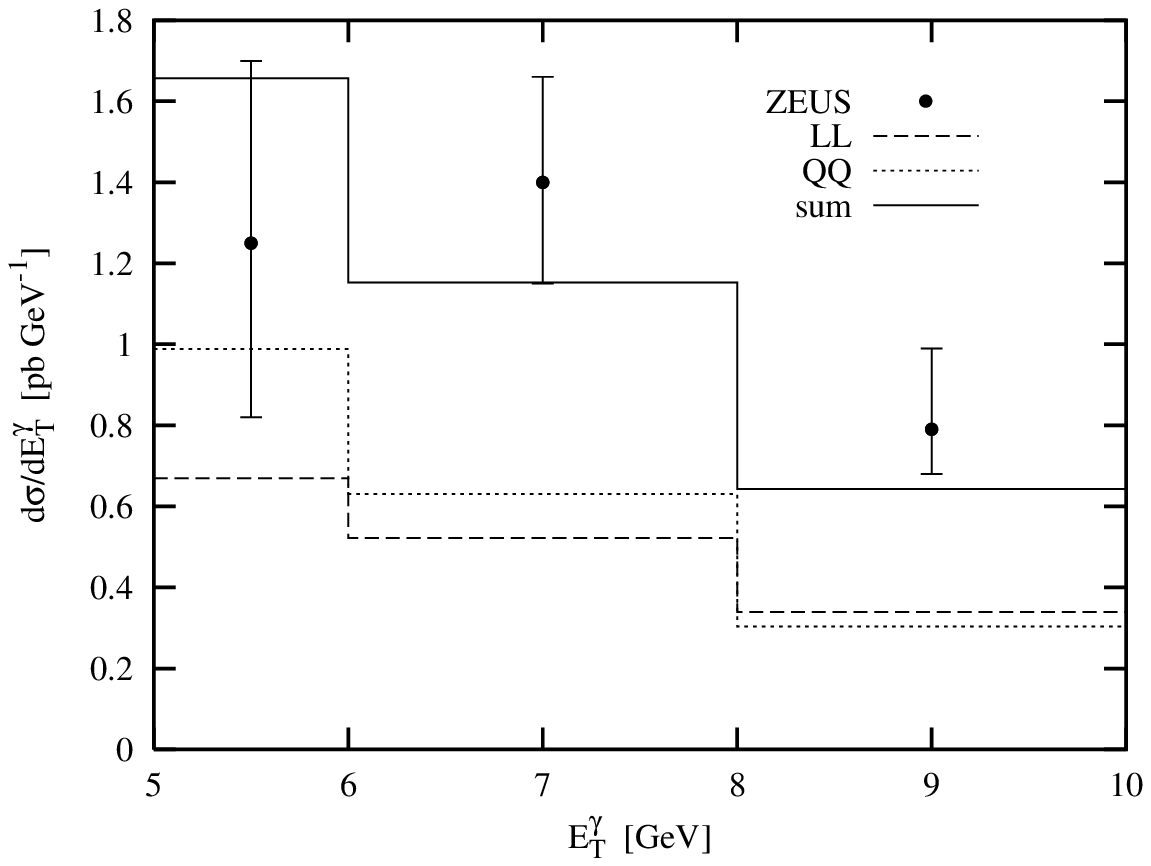,angle=-90,width=6.1cm} 
\caption{Transverse momentum distribution of isolated photons, 
compared to ZEUS data.\label{fig:et}}
\epsfig{file=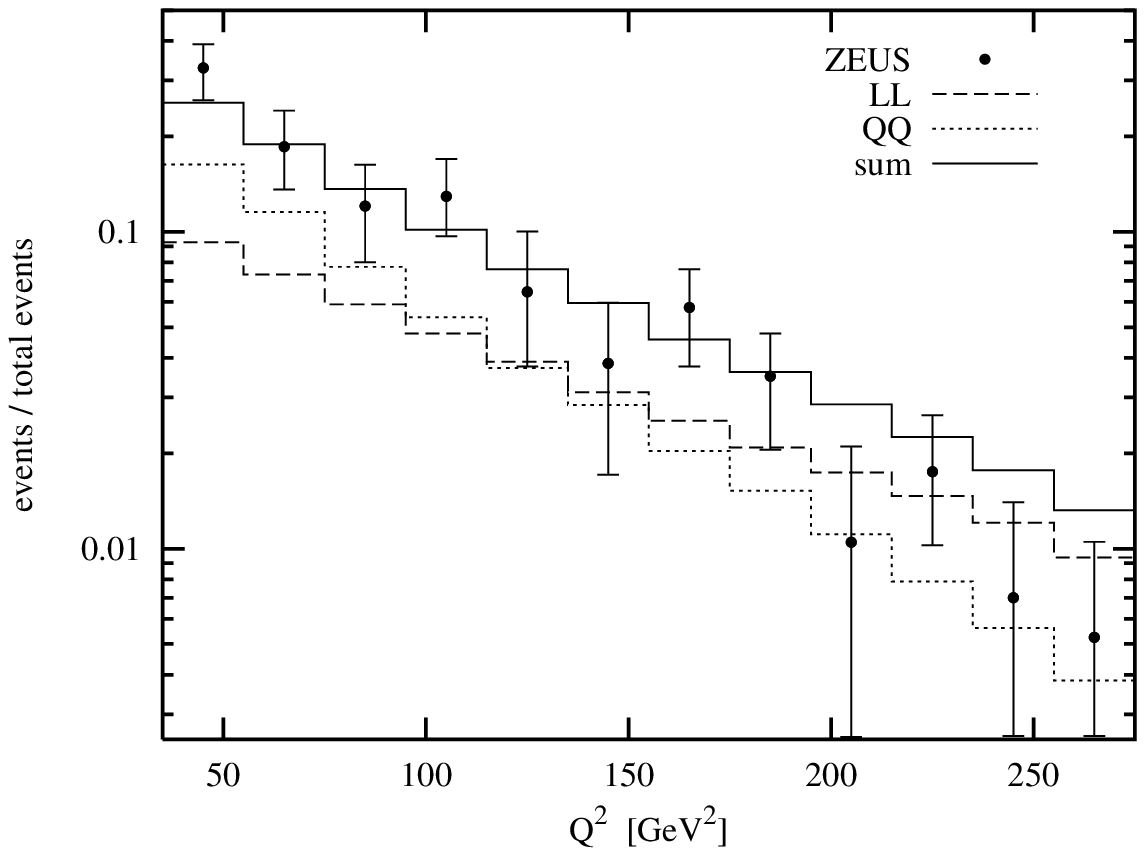,angle=-90,width=6.3cm} 
\caption{Dependence of isolated photon cross section on 
$Q^2$, as reconstructed from the outgoing electron. Data and theory are 
normalized to the total number of events, as in the experimental 
ZEUS measurement.
\label{fig:qsq}}
\end{figure}

We also computed the individual contributions to this cross 
section, which we list in Table~\ref{tab_werte}. It can be seen that 
the difference among the different beam energies and $e^+/e^-$ induced 
cross sections is about 6\%, thus justifying their combination into a single 
data sample. By decomposing the observed cross section into the $QQ$, $LL$ and 
$QL$ contributions, we find that the $QQ$ contribution yields only 
53\% of the cross section, although the experimental cuts were designed to 
enhance this contribution relative to the others. Especially, by 
requiring the final state lepton and the photon to be found in different
parts of the detector, any small-angle radiation off the lepton is 
suppressed, thus leaving only the (kinematically disfavored) 
large-angle radiation in the $LL$ contribution. 
The still substantial magnitude of the $LL$ contribution can be understood by 
the larger magnitude of the electric charge of the lepton compared with the
quark. As expected, the $QL$ contribution is very small. Finally, we 
observe that using the BFG (type I) parameterization~\cite{bfg}
 for $D_{q\to \gamma}(z)$ instead of the ALEPH parameterization~\cite{aleph}
enhances the theoretical prediction only insignificantly
by two per cent.

To investigate the dependence of the isolated photon cross section on the 
event kinematics, the ZEUS collaboration also measured differential 
distributions in $\eta_\gamma$, $E_{T,\gamma}$ and $Q^2$. Comparing the 
shapes of these distributions with normalized predictions from
HERWIG~\cite{herwig} and 
PYTHIA~\cite{pythia}, it was found that 
none of these programs could describe all distributions:  both 
reproduced only the  shape of the  $E_{T,\gamma}$-distribution 
correctly,
HERWIG  predicted a too soft $Q^2$-distribution, while PYTHIA yielded 
an incorrect $\eta_\gamma$-distribution.
The approach suggested in~\cite{mrst}, containing only the $LL$ subprocess,
was found to yield a reasonable description of the 
$E_{T,\gamma}$-distribution, but failed on the 
$\eta_\gamma$-distribution~\cite{saxon}.
Using our leading order calculation, we obtain differential cross sections 
in $\eta_\gamma$, $E_{T,\gamma}$ and $Q^2$, which are shown in
 Figures~\ref{fig:rap},~\ref{fig:et} and \ref{fig:qsq}. Comparison with the 
ZEUS data shows agreement for all three distributions in both shape and 
normalization. It should be noted that ZEUS does not provide a differential 
distribution in $Q^2$, but just normalized event counts binned in this 
variable.

Especially the $\eta_\gamma$-distribution, Figure~\ref{fig:rap}, 
gives important insight into the discrepancies observed between the data 
and predictions from PYTHIA and HERWIG. In this distribution, the shapes of 
the $QQ$ and $LL$ contributions are considerably different.
Comparing with the distributions obtained in~\cite{zeus} from the 
event generator programs, it can be seen that the shape 
of the $QQ$ contribution resembles the PYTHIA prediction, while the 
shape of the $LL$ contribution resembles the HERWIG prediction. This
observation suggests that each program accounts for only one 
of the  subprocesses 
 appropriately: PYTHIA for only $QQ$ and HERWIG only for $LL$. 
The lack of photon radiation off quarks in HERWIG was 
already observed by the H1 collaboration~\cite{kmuller} in the study of 
 photoproduction of isolated photons. 
The importance of both subprocesses for the 
shape of the $\eta_\gamma$-distribution
shows clearly that the 
isolated photon cross section in DIS can not  be described by the 
$LL$ subprocess~\cite{mrst} only. 

In this letter, we investigated the production of isolated photons in 
deep inelastic scattering in view of a recent ZEUS measurement of this 
observable. We found that 
photon radiation off quarks and leptons contribute 
about equal amounts to this observable, although radiation off leptons is 
restricted to large angles by the kinematical cuts. 
Since the photon 
isolation criterion 
admits some amount of hadronic activity around the photon direction, 
small angle radiation off quarks is kinematically allowed, and 
inherently contains a contribution from the 
non-perturbative quark-to-photon fragmentation function. 
Both these 
effects (large-angle radiation and photon fragmentation) are included in
our fixed-order parton model calculation, 
while they are usually not accounted for in standard event generator 
programs. While the ZEUS collaboration could not describe their data 
with event generator predictions, we found that our calculation is in 
very good agreement with the ZEUS data both in normalization and in shape.

{\bf Acknowledgment:} 
We would like to thank Katharina M\"uller, Carsten Schmitz,  Ueli Straumann
and David Saxon
for many useful and clarifying discussions. 
This work was supported by the Swiss National Science Foundation 
(SNF) under contract PMPD2-106101.


\end{document}